\documentclass[12pt]{article}
\pdfoutput=1
\usepackage{amsfonts,amssymb,epsfig,amsmath}
\usepackage{color}

\usepackage[vcentermath]{youngtab}
\renewcommand{\baselinestretch}{1.2}

\begin{document}

\makeatletter \@addtoreset{equation}{section} \makeatother
\renewcommand{\theequation}{\thesection.\arabic{equation}}
\renewcommand{\thefootnote}{\alph{footnote}}

\begin{titlepage}

\begin{center}
\hfill {\tt SNUTP11-010}\\
\hfill {\tt KIAS-P12011}\\
%\hfill {\tt arXiv:11mm.nnnn}

\vspace{2.5cm}

{\large\bf Tests of AdS$_4$/CFT$_3$ correspondence %via index computation
for $\mathcal{N}=2$ chiral-like theory}

\vspace{2cm}

\renewcommand{\thefootnote}{\alph{footnote}}

{%\large
 Dongmin Gang$^1$, Chiung Hwang $^2$, Seok Kim$^3$ and Jaemo Park$^2$}

\vspace{1cm}

\textit{$^1$School of Physics, Korea Institute for Advanced Study, Seoul
130-722, Korea.}

%\textit{$^1$School of Physics, Korea Institute for Advanced Study, Seoul
%130-012, Korea.}\\

\vspace{0.2cm}

\textit{ $^2$Department of Physics \& Center for Theoretical Physics (PCTP),\\
POSTECH, Pohang 790-784, Korea.}

\vspace{0.2cm}

 \textit{$^3$Department of Physics and Astronomy \&
Center for
Theoretical Physics,\\
Seoul National University, Seoul 151-747, Korea.}\\

\vspace{0.7cm}

%E-mails: {\tt skim@phya.snu.ac.kr, }

\end{center}

\vspace{2cm}

\begin{abstract}

We investigate the superconformal index and the partition function
for the chiral-like Chern-Simons-matter theory proposed for
M2-branes probing the cones over $M^{3,2}$ and find perfect agreements
with the gravity index and the gravitational free energy.

\end{abstract}

\end{titlepage}

\renewcommand{\thefootnote}{\arabic{footnote}}

\setcounter{footnote}{0}

\renewcommand{\baselinestretch}{1}

\tableofcontents

\renewcommand{\baselinestretch}{1.2}

\section{Introduction}

There has been rapid progress in our understanding of $AdS_4/CFT_3$
correspondence in recent years. Given $AdS_4 \times Y$ with $Y$
being a Sasaki-Einstein 7-manifold, the corresponding superconformal
field theory is realized as a supersymmetric Chern-Simons matter
(SCSM) theory \cite{sch}. The general theories with $\mathcal{N}\geq
4$ supersymmetry were constructed in \cite{HLLLP, HLLLP2} and a
famous example out of such construction is the ABJM theory
\cite{ABJM}, which describes M2 branes on $C^4/Z_k$ with $k$ being
related to the Chern-Simons level of the field theory. There have
been various checks made for theories with $\mathcal{N}\geq 3$
supersymmetry such as the partition function, the famous $N^{3/2}$
behavior and the superconformal index \cite{Drukker, Herzog:2010hf,
Kim:2009wb, Imamura:2009hc, Imamura:2010sa,
Sangmin08,Imamura:2011su,Imamura:2011uj}. Some other aspects of the
superconformal index are explored in , e.g., \cite{Vartanov,
Willett}.

If we consider theories with $\mathcal{N}=2$ supersymmetry, we have
much more diverse possibilities. It is not clear at this moment how
to find field theory dual to a given gravitational background $AdS_4
\times Y$ with $\mathcal{N}=2$ supersymmetry. For some special cases
such as $M^{3,2}, Q^{1,1,1}, V^{5,2}, Y^{p,q}$, there are various
proposals on the field theory duals
\cite{Jafferis:2009th,Benini,Franco:2008um,Martelli:2008si,
Hanany:2008cd, Benini:2011cma, Martelli09}. A subtlety is that
R-charges of the matters in such theories can have different values
from the canonical one. It is proposed in \cite{Jafferis2} that the
R-charge can be determined by minimizing the partition function of
the field theory defined on $S^3$. Given this description, one can
compute the partition function on $S^3$ of the proposed field theory
and compare it with the gravitational free energy. Various
impressive results are reported \cite{Jafferis:2011zi, Cheon:2011vi,
Martelli11-2}. Also, for a special class of theories such as
$M^{3,2}, Q^{1,1,1}$, the superconformal index is computed and
comparison is made with the gravity index \cite{Cheon:2011th}.

A curious technical aspect in computing the index and
the partition function is that it is much easier to work out
non-chiral theories in 4d sense. We call 3d theories `chiral-like
theories' if the matter fields are inherited from chiral matters in 4d.
These chiral-like theories could suffer from parity anomalies and the
right index computation needs to take care of this issue properly. In the
large $N$ computation of the partition function on $S^3$, one
employs the saddle point approximation and it was crucial to have
the vanishing long range force among the eigenvalues of the matrix
model. This condition was not satisfied for chiral-like theories.
However it is reasonable to expect that this obstacle is  simply a
technical one. The purpose of the current paper is to show that
chiral-like theories can be dual to $AdS_4$ gravitational
backgrounds. Specifically we consider one chiral-like theory, which
is expected to be dual to $AdS_4 \times M^{3,2}$ (with nonzero discrete torsions).
We compute its superconformal index and the partition function on $S^3$ to find
perfect agreements with the gravity index and the gravitational free
energy, respectively.

In \cite{Cheon:2011th}, the index computation was carried out for
the same theory as we considered here. Facing the parity anomaly,
we have only considered the topological sectors in which the path integral
is well defined.
%different Chern-Simons level of $U(1)$ factors from $SU(N)$ level is
%assigned to make the theory consistent.
It was observed that this
field theory index contains the dual gravity spectrum. However it
also contains additional contribution, whose physical meaning was not
clear. In \cite{Benini:2011cma}, a method was proposed to
resolve the parity anomaly. It turns out that if we introduce the
off-diagonal Chern-Simons terms for $U(1)$ factors of $U(N)$ gauge
groups, the theory becomes consistent. In this paper, we compute the
superconformal index in the presence of the proposed off-diagonal
Chern-Simons terms to find a perfect agreement with the
gravity spectrum.

For the computation of the partition function, one of the crucial
points of usual saddle point approximation of non- chiral  theories
is the  cancelation of the long range forces on eigenvalues of the
matrix model. Chiral-like theories do not enjoy this feature. In
\cite{Amariti:2011jp}, it is suggested that if we consider the
symmetrized partition function, it is possible that chiral-like
theories can have the cancelation of the long range force. The
underlying idea is a very simple one. If we consider the partition
function, it has the form
\begin{equation}
Z=\int du F(u)
\end{equation}
where the integration is done over the Cartan elements of the gauge
group. As the measure $F(u)$ is not invariant under a $Z_2$
transformation $u\rightarrow -u$, the saddle point approximation does not
exhibit this symmetry either. The proposal of \cite{Amariti:2011jp} is to
consider the symmetrized measure
\begin{equation}
Z=\frac{1}{2} \int du (F(u)+F(-u))
\end{equation}
which gives the same $Z$. Then the integrand of the integral has the
$Z_2$ symmetry and the saddle point equation also respects this symmetry.
We employ this method to the $M^{3,2}$ model to find that (log of) the
partition function nicely agrees with the gravitational free energy.

The content of the paper is as follows. After the introduction, we
introduce the field theory model of $M^{3,2}$ and carry out the
index computation in section 2. In section 3, we work out the
partition function on $S^3$ of the field theory to match the
gravitational free energy.

As this work is completed, we receive the paper by Amariti et. al.
\cite{Amariti11}, where similar topic is covered. However they just
consider the partition function and we also consider the
superconformal index for $M^{3,2}$ model to add more impressive
evidences for $AdS_4/CFT_3$ correspondence.

\section{Index computation}

\subsection{Review of previous results}

\begin{figure}[h!]
  \begin{center}
    \includegraphics[width=6cm]{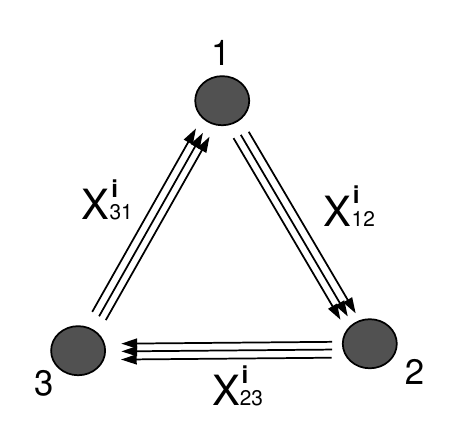}
\caption{The quiver diagram of M2-branes on
$C(M^{3,2})$}\label{quiver-m32}
  \end{center}
\end{figure}

The field theory proposed for M2 branes on $C(M^{3,2})$ is given by
the quiver diagram in Fig \ref{quiver-m32}. The Chern-Simons level
for $U(N)$ is given by $(-2k,k,k)$ for $M^{3,2}/Z_k$.  For the
brevity of notation, we shall often denote the fields by
$X^I_{23}=A^I$, $X^I_{31}=B^I$, $X^I_{12}=C^I$. $I=1,2,3$ is the
triplet index for the $SU(3)$ global symmetry.
This theory was initially studied by \cite{Martelli:2008si,
Hanany:2008cd} and it was recently argued at \cite{Benini:2011cma}
that this model is dual to $AdS_4\times M^{3,2}$ with discrete
torsion. Since the presence of the discrete torsion does not affect
the gravity index in the large N limit, one can try to match the
field theory index to that of the gravity side. The gravity index is
given by the Plesythetic form
\begin{equation}
I(x,y_1,y_2,y_3)=exp(\sum_n \frac{1}{n}
I_{sp}(x^n,y^n_1,y^n_2,y^n_3))
\end{equation}
where $x$, $y_3$ are fugacities of the `energy+angular momentum' and the Cartan
of an $SU(2)$ isometry of $M^{3,2}$, respectively. $y_1,y_2$ are the fugacities of
the $SU(3)$ Cartans. See \cite{Cheon:2011th} for more details. The $SU(2)$ Cartan
is mapped to the $U(1)_B$ symmetry carried by the sum of monopole charges of the
first gauge group $U(N)_{-2}$ where the subscript denotes the Chern-Simons level.
For instance, at $y_1=y_2=1$, the single particle
index is given by
\begin{equation}
I_{sp}(x,y_3)=\frac{9x^2y_3}{(1-x^2y_3)^2}+\frac{9x^2y_3^{-1}}{(1-x^2y_3^{-1})^2}.
\end{equation}
From the form of the single particle index, one can see that the
large $N$ index exhibit the factorization structure
\begin{eqnarray}
I(x,y_1,y_2,y_3)= I_{0}(x,y_1,y_2)\hat{I}(x,y_1,y_2,y_3)\;.
\end{eqnarray}
Here $I_{0}$ is the multi-particle gravity index from $U(1)_B$
neutral gravitons, which should match the large $N$ index of
the field theory in the zero monopole flux sector. $\hat{I}$ is the
multi particle gravity index from gravitons carrying nonzero $U(1)_B$
charges. In
the field theory side,  $\hat{I}$ should be obtained by summing all
large $N$ index contribution $I^\prime_{(p,q,r)}$ from non-zero
monopole charges $\{p_i\},\{q_i \}, \{r_i\}$ in $U(N)_{-2}, U(N)_1,
U(N)_1$ gauge groups. One subtlety of the field theory of interest
is that it is chiral in the 4d sense so it has odd number of
bifundamental fields for each gauge group, hence it suffers from parity
anomaly. It is shown in \cite{Cheon:2011th} that this parity anomaly
makes the path integral inconsistent for certain values of monopole
fluxes in the overall $U(1)$ of each $U(N)$ factor. \cite{Cheon:2011th}
simply considered the sectors which are free of these anomaly effects.
%this parity anomaly
%could affects the flux quantization condition for overall $U(1)$ of
%each $U(N)$ factor.
%In \cite{Cheon:2011th}, different quantization
%condition for Chern-Simons level for $U(1)$ of $U(N)$ is imposed to
%make the theory consistent.

Schematically, the large $N$ field theory index $I$ is given by
\begin{eqnarray}
\hspace*{-1cm}I&=&\frac{x^{\epsilon_0 }}{({\rm symmetry})}\int\left[
  \frac{d\alpha d\beta d\gamma}{(2\pi)^3}\right]e^{i S^{(0)}_{CS} ( \alpha, \beta, \gamma)}
  e^{ib_0(\alpha,\beta,\gamma)} \exp\left[\sum_{n=1}^\infty\frac{1}{n}  f_{sp} (\cdot^n) \right] \; . \label{large N index}
  \end{eqnarray}
where the integration is done over the holonomy variables $\alpha_i,
\beta_i, \gamma_i, \,\,\, i=1 \cdots N$ of three $U(N)$s,
$\epsilon_0$ is the zero point energy, $S^{(0)}_{CS}$ is the action,
$b_0$ is the zero point charge contribution for a suitable saddle
point configuration, which is specified by the monopole charges
$\{p_i\},\{q_i \}, \{r_i\}$ in $U(N)_{-2}, U(N)_1, U(N)_1$ gauge
groups. $f_{sp}$ is the single particle or the `letter' index.
Explicit expressions for $\epsilon_0 , S^{(0)}_{CS}, b_0$ and
$f^\prime (\cdot^n)$ and the explanation of other variables are
given in \cite{Cheon:2011th}.\footnote{Note that the large $N$ field
theory index $I(x)$ is obtained by holonomy integration over the
integrand containing $f_{sp}$. Similarly $f^\prime (\cdot^n)$
denotes the single particle index with non-zero monopole flux, which
gives rise to $I^\prime (x)$ in eq. (\ref{Iprime}) after holonomy
integration with summing over all non-zero monopole fluxes.} In the
index formula, we take the dimensions of three chiral fields $A^I,
B^I, C^I$ to be $a,b,c$ respectively . Taking into account of
marginality of superpotential and discrete symmetry in the quiver
diagram, one can take
 \begin{eqnarray}
 a = 2-2b, \quad c=b \; .
 \end{eqnarray}
It is shown  in \cite{Cheon:2011th} that the zero monopole sector
perfectly agrees with $I_0$ of the gravity index. The large $N$  index
with nonzero monopole sectors $I^\prime (x)$ can be divided into two pieces,
 \begin{eqnarray}
 I^\prime (x) =  \hat{I} (x) +I^{b} (x) \; .  \label{Iprime}
 \end{eqnarray}
$\hat{I}$ is the large $N$ index from factorizable monopole fluxes and
$I^b$ is from non-factorizable fluxes. Factorizable fluxes satisfy
$P_\pm = Q_\pm  = R_\pm$ where $P_+=\sum p_{i+}, P_-=\sum p_{i-}$ denote
the sums of positive/negative fluxes in $\{p_i\}$ and $Q_\pm, R_\pm$ are defined
in a similar way. Factorizable fluxes have the following properties
which distinguish them from non-factorizable fluxes.
\\
\\
1. Factorization:   $I_{(p, q, r )} = I_{(p_+ , q_+, r_+ )}I_{(p_{-} ,
q_{-}, r_{-} )}$\;.
\\
2. $I_{(p,q,r)}$ is independent of R-charge assignment $b$ that we have not
specified.
\\
\\
Here $p_\pm, q_\pm , r_\pm$ denote the positive/negative fluxes in
$\{p_i \}, \{q_i\}$ and $\{r_i\}$ respectively. In the $\hat{I}$ part of
the index, each factor with definite nonzero value of $P_+$ or $P_-$ gives
the same spectrum as the gravity index coming from multi-gravitons carrying
positive or negative $U(1)_B$ charges $\pm P_\pm$, at least for a
few low order terms in $x$. $I^b$ part gave additional spectrum
which does not appear in the gravity side.

\subsection{Index computation with mixed Chern-Simons terms}

In \cite{Benini:2011cma}, it was proposed to resolve the parity
anomaly by introducing off-diagonal Chern-Simons terms for  three
$U(1)$s in $U(N)$s. The off diagonal CS term is given by
\begin{eqnarray}
\mathcal{L}_{\textrm{off-CS}} =  \sum_{a,b=1}^3
\frac{\Lambda_{ab}}{4 \pi} \int \textrm{Tr} A_a \wedge d
\textrm{Tr}A_b \; .
\end{eqnarray}
$\Lambda_{ab}$ are symmetric $3\times 3$ matrices satisfying the
following conditions
\begin{eqnarray}
&&\Lambda_{ab} - \frac{1}{2} A_{ab} \in \mathbb{Z} \quad \quad
\forall  a, b \; . \label{off CS cond1}
\\
&& \sum_{a=1}^3 \Lambda_{ab} =0  \quad \quad b=1,2,3,\; . \label{off
CS cond2}
\end{eqnarray}
where $A_{ab} $ are adjacency matrix for the quiver diagram,
$A_{12}\!=\!A_{23}\!=\!A_{31}\!=\!3$. The first condition
comes from the parity anomaly cancellation conditions and the second
guarantees that chiral ring structure is not modified  after
introducing  off diagonal CS terms. Let us call the diagonal
components of $\Lambda$ as $l_1,l_2$ and $l_3$, which determine
the off-diagonal elements from \eqref{off CS cond2}:
\begin{eqnarray}
\Lambda=\left(\begin{array}{ccc}
l_1&\frac{1}{2}(-l_1-l_2+l_3)&\frac{1}{2}(-l_1+l_2-l_3)\\
\frac{1}{2}(-l_1-l_2+l_3)&l_2&\frac{1}{2}(l_1-l_2-l_3)\\
\frac{1}{2}(-l_1+l_2-l_3)&\frac{1}{2}(l_1-l_2-l_3)&l_3
\end{array}\right) \;.
\end{eqnarray}
From the conditions \eqref{off CS cond1}, $l_1, l_2, l_3$ should be
integers satisfying the condition that $\frac{1}{2}(-l_1-l_2+l_3)$ is
half of an odd integer. As in \cite{Cheon:2011th}, for generic monopole
flux one obtains states whose energies are of order $\mathcal{O}(N)$. As we
are here interested in comparing our index with the low energy gravity index
at large $N$, let us analyze the conditions for monopole fluxes which yield
$\mathcal O(1)$ energy contribution in the presence of off-diagonal CS terms.
With the CS levels $k=(-2,1,1)$, holonomies $(\alpha,\beta,\gamma)$ and the
magnetic fluxes $(p,q,r)$, the phase factors in the index are given by
\begin{eqnarray}
S^{(0)}_{CS}&=&\sum_i(-2p_i\alpha_i+q_i\beta_i+r_i\gamma_i)+\sum_{i,j}(l_1p_i\alpha_j+l_2q_i\beta_j+l_3r_i\gamma_j)\nonumber\\
&&+\frac{1}{2}\sum_{i,j}\bigl[(-l_1-l_2+l_3)(p_i\beta_j+q_i\alpha_j)+(l_1-l_2-l_3)(q_i\gamma_j+r_i\beta_j)
+(-l_1+l_2-l_3)(r_i\alpha_j+p_i\gamma_j)\bigr],\nonumber\\
\\
b_0&=&\frac{3}{2}\left[\sum_{i,j}|p_i-q_j|(\alpha_i-\beta_j)+\sum_{i,j}|q_i-r_j|(\beta_i-\gamma_j)+\sum_{i,j}|r_i-p_j|
(\gamma_i-\alpha_j)\right].
\end{eqnarray}
Let us pick a holonomy $\alpha_i$ whose corresponding flux $p_i$ is
zero. The phase term containing this $\alpha_i$ is given by
\begin{equation}\label{coefficient:holonomy}
\left[l_1\sum_jp_j+\frac{1}{2}(-l_1-l_2+l_3)\sum_jq_j+\frac{1}{2}(-l_1+l_2-l_3)\sum_jr_j
+\frac{3}{2}\left(\sum_j|q_j|-\sum_j|r_j|\right)\right]\alpha_i\ .
\end{equation}
Since we are looking for states of  $\mathcal O(1)$ energy in the
large $N$ limit, only $\mathcal O(1)$ number of holonomies should
survive in the phase factor, as all phases have to be canceled by
exciting matter fields which carry nonzero energies. This means that
the coefficient in \eqref{coefficient:holonomy} must be zero. This argument
also applies to both $\beta$ and $\gamma$ whose corresponding magnetic fluxes
are zero. We therefore obtain the following equations
\begin{eqnarray}
&&\left(\begin{array}{c}
\Delta_1\\
\Delta_2\\
\Delta_3
\end{array}\right)\equiv\left(\begin{array}{ccc}
0&1&-1\\
-1&0&1\\
1&-1&0
\end{array}\right)\left(\begin{array}{c}
|P|\\
|Q|\\
|R|
\end{array}\right),\label{condition:|p|}\\
&&\left(\begin{array}{ccc}
l_1&\frac{1}{2}(-l_1-l_2+l_3)&\frac{1}{2}(-l_1+l_2-l_3)\\
\frac{1}{2}(-l_1-l_2+l_3)&l_2&\frac{1}{2}(l_1-l_2-l_3)\\
\frac{1}{2}(-l_1+l_2-l_3)&\frac{1}{2}(l_1-l_2-l_3)&l_3
\end{array}\right)\left(\begin{array}{c}
P\\
Q\\
R
\end{array}\right)=-\frac{3}{2}\left(\begin{array}{c}
\Delta_1\\
\Delta_2\\
\Delta_3
\end{array}\right).\label{condition:p}\nonumber\\
\end{eqnarray}
where $|P|=\sum_j|p_j| = P_+ - P_-$ and $P=\sum_jp_j = P_+ + P_-$.
Note that $\sum_j\Delta_j=0$. Let us also consider the zero point energy $\epsilon_0$,
\begin{eqnarray}
\epsilon_0&=&\frac{3}{2}\left((1-c)\sum_{i,j}|p_i-q_j|+(1-a)\sum_{i,j}|q_i-r_j|+(1-b)\sum_{i,j}|r_i-p_j|\right)\nonumber\\
&&-\sum_{i<j}(|p_i-p_j|+|q_i-q_j|+|r_i-r_j|)\ ,
\end{eqnarray}
which could also yield an $\mathcal{O}(N)$ energy contribution.
The possible $\mathcal O(N)$ contribution is \cite{Cheon:2011th}
\begin{equation}
N_1\left(\frac{3}{2}(1-c)|Q|+\frac{3}{2}(1-b)|R|-|P|\right)+cyclic
\end{equation}
where $N_1$ denotes the number $U(1)$'s in the Cartan of first $U(N)$
which support zero magnetic fluxes. By substituting $|R|=|P|+\Delta_2$ and
$|Q|=|P|-\Delta_3$  obtained from \eqref{condition:|p|}, we can check after
some algebra that $\Delta_2$ and $\Delta_3$ must be the same to prevent the
$\mathcal{O}(N)$ contribution from $\epsilon_0$. Thus, \eqref{condition:p} can
be solved as
\begin{eqnarray}
Q&=&P+\frac{3\left(-l_1+l_2+3 l_3\right)}{l_1^2+\left(l_2-l_3\right){}^2-2 l_1 \left(l_2+l_3\right)}\Delta\label{q}\\
R&=&P+\frac{3\left(-l_1+3
l_2+l_3\right)}{l_1^2+\left(l_2-l_3\right){}^2-2 l_1
\left(l_2+l_3\right)}\Delta\label{r}
\end{eqnarray}
where $\Delta\equiv\Delta_2=\Delta_3$. As the diagonal combination of
$U(1)\subset U(N)_{-2}\times U(N)_1\times U(N)_1$ decouples from
matter fields, the fluxes should satisfy the condition
\begin{equation}\label{condition:diagonal}
2 P=Q+R,
\end{equation}
which is unaffected by $\Lambda$ due to the condition \eqref{off CS
cond1}. Thus, if we choose $\Lambda$ such that
\begin{equation}\label{off CS cond3}
  -l_1+2l_2+2l_3\neq0\ ,
\end{equation}
then $\Delta$ must be zero.\footnote{(\ref{off CS cond3}) is the only extra condition
on the off-diagonal Chern-Simons term that we claim should be imposed, apart from
(\ref{off CS cond1}) and (\ref{off CS cond2}) imposed by \cite{Benini:2011cma}.}
In this case, $|P|=|Q|=|R|$ and $P=Q=R$, which implies that
\begin{equation}
P_\pm=Q_\pm=R_\pm \;.
\end{equation}
This is nothing but the condition for factorizable fluxes. Since
$I^b$ comes from non-factorizable fluxes, it does not appear in the
large $N$ index if we choose off diagonal CS term satisfying
(\ref{off CS cond3}). Furthermore, for factorizable fluxes, one can
immediately check using (\ref{off CS cond2}) that the additional CS
term with the levels $\Lambda_{ij}$ vanishes, which means that the
index contribution is not changed by this additional CS term. Thus,
the large $N$ index with (\ref{off CS cond3}) gives rise to
\begin{eqnarray}
I^\prime (x) = \hat{I}(x)\ .
\end{eqnarray}
This monopole index is shown to agree with the gravity index
\cite{Cheon:2011th}. Furthermore, now we can see the explicit
decomposition of the index into $U(1)_B$ neutral, positive and
negative sectors. We already saw that $O(1)$ contribution to the
large $N$ index is independent of the various choices $l_1,l_2,l_3$.
However, various choices of $l_1,l_2,l_3$ possibly give different
results for $O(N)$ contribution. Thus if we are interested in the
baryon spectrum, this subtlety can play an important role. It is
worthwhile to explore this issue.

So far we consider the so-called $CP$ non-invariant model for
$M^{3,2}$, which is claimed to be dual to $M$-theory on $AdS_4
\times M^{3,2}$ with discrete torsions. In \cite{Benini:2011cma}, CP
invariant field theory model dual to $M^{3,2}$ without discrete
torsion is proposed as well. The theory is given by $U(N-2)_0 \times
U(N)_0\times U(N)_0$ CS theory with the matter contents given by the
same quiver diagram as the CP non-invariant case. We attempt to
calculate the large $N$ superconformal index for the theory and
compare it with the gravity spectrum. However, it is hard to see the
index matching in this case. Although the $U(1)_B$ neutral sector
$I_0$ is the same as that of $U(N)_{-2}\times U(N)_1\times U(N)_1$
theory, which matches well with the gravity index, the sector with
$U(1)_B\textrm{ charge}=1$ with $P_+=1$ and $P_-=0$ is not. For
$p=q=r=1$, we find $b_0=3(\beta-\gamma)$ and $\epsilon_0=2-3a$. The
lowest energy contribution appears by exciting matters
$A_{i}A_{j}A_{k}$ to screen the phase from monopoles, which have
$\epsilon+j=3a+(2-3a)=2$. As $SU(3)$ indices $i,j,k$ are
symmetrized, these form a $10$ dimensional representation
contributing to the index as $10x^2$. This is different from the
gravity index $I^{grav}_{P_+=1}=9x^2$. Other $U(1)_B\textrm{
charge}=1$ sectors with $P_+=n+1$ and $P_-=n$ do not help because
they seem to yield contributions to higher orders than $x^2$. We
checked this by studying $b_0$ and $\epsilon_0$ in these sectors. If
we check higher $U(1)_B$ charge sectors, the situation is worse. For
$P_+=n\geq2$, monopole fluxes of the form
\begin{equation}
  p=(2,\underbrace{1,\cdots,1}_{n-2}, 0, \cdots)\ ,\ \ q=(\underbrace{1,\cdots,1}_{n},
  0, \cdots)\ ,\ \ r=(\underbrace{1,\cdots,1}_{n}, 0, \cdots)
\end{equation}
all have $b_0=0$ and $\epsilon_0=2$. As there are no phases created by monopoles,
they are gauge invariant without exciting matters. So they start to come with
energy $2$, contributing to the index as $x^2+\cdots$. This means that there are
infinite number of states at $\epsilon+j=2$. This kind of divergence is similar to that
observed in the index computation of $\mathcal{N}=8$ super Yang-Mills
\cite{ABCD}. At this point, it is not clear how to cure this problem
in the index computation.

\section{$S^3$ partition function and the gravitational free energy}

 The gravitational free energy of $AdS_4 \times Y$ with $Y$
being Sasaki-Einstein 7 manifold is given by
\begin{equation}
F_{grav}=N^\frac{3}{2}\sqrt{\frac{2\pi^6}{27\textrm{Vol}(Y)}}.
\end{equation}
in the leading order of $N$ with the normalization
$R_{\hat\mu\hat\nu}=6g_{\hat\mu\hat\nu}$ of the metric on $Y$. The
volume of $M^{32}$ is given by \cite{Fabbri:1999hw}
\begin{equation}
\textrm{Vol}(M^{3,2})=\frac{9\pi^4}{128}.
\end{equation}
If we consider $M^{3,2}/\mathbb{Z}_k$, the volume of the space is
divided by $k$ so that the free energy is given by\footnote{There
could be a subtlety in the formula since $M^{3,2}/\mathbb{Z}_k$ can
have orbifold singularities. However, the leading large $N$ results
are expected not to be affected by the orbifold singularities.}
\begin{equation}
F_{grav}=\frac{16\sqrt3\pi}{27}k^{\frac{1}{2}}N^\frac{3}{2}.
\end{equation}

The localization computes the partition function of the field theory
\begin{eqnarray}\label{partition function:original expression}
Z&=&\int\left(\prod_\textrm{Cartan}\frac{d\sigma}{2\pi}\right)\exp\left[\frac{ik}{4\pi}
\sum_i\left(-2{\sigma_{1,i}}^2+{\sigma_{2,i}}^2+{\sigma_{3,i}}^2\right)+\frac{i}{4\pi}
\sum_{a,b=1}^3\sum_{i,j}\Lambda_{ab}\sigma_{a,i}\sigma_{b,j}-\sum_{a=1}^3\sum_i\Delta^m_a
\sigma_{a,i}\right]\nonumber\\
&&\left(\prod_{a=1}^3\prod_{i,j,i\neq
j}2\sinh\frac{\sigma_{a,i}-\sigma_{a,j}}{2}\right)
\exp\left[\sum_I\sum_{i,j}\ell\left(1-\Delta^I_{12}+i\frac{\sigma_{1,i}-\sigma_{2,j}}
{2\pi}\right)+cyclic\right].\nonumber\\
\end{eqnarray}
where $\Delta^m_a$ is the R-charge for a bare monopole operator
corresponding to a unit flux of ${\rm tr}F_a$
\cite{Jafferis:2011zi}, $\Delta^I_{ab}$ is the R-charge of the
bifundamental field $X^I_{ab}$ and $\sigma$s are the scalars in the
vector multiplets. The function $\ell(z)$ is given by
\begin{equation}
\ell(z)=-z\ln\left(1-e^{2\pi iz}\right)+\frac{i}{2}\left(\pi
z^2+\frac{1}{\pi}\textrm{Li}_2\left(e^{2\pi
iz}\right)\right)-\frac{i\pi}{12}.
\end{equation}
 Let us check the obvious
flat directions of the partition function, whose consideration is
crucial to the later calculation. We can see that the partition
function is invariant under the following transformations up to a
phase:
\begin{eqnarray}\label{flat directions}
\sigma_{a}&\rightarrow&\sigma_{a}-2\pi i\delta_a,\nonumber\\
\Delta^I_{ab}&\rightarrow&\Delta^I_{ab}+\delta_a-\delta_b,\\
\Delta^m_a&\rightarrow&\Delta^m_a+k_a\delta_a+N\sum_b\Lambda_{ab}\delta_b\nonumber
\end{eqnarray}
with 3 parameters $\delta_a$. Due to this symmetry, we can adjust the
R-charge of the bifundamental fields as follows:
\begin{equation}
\Delta^I_{ab}\rightarrow\Delta^I=\frac{\Delta^I_{12}+\Delta^I_{23}+\Delta^I_{31}}{3}.
\label{DeltaI}
\end{equation}
The marginality of the superpotential demands that
$\sum_I\Delta^I=2$.

One important point is that, as we consider the $Z_2$ transformation
of the integration variables $\sigma_i\rightarrow -\sigma_i$, the integral
acquires nonzero contribution only from the $Z_2$ even part of the integrand.
We want to evaluate this partition function in the large $N$ limit. In this
limit, we can evaluate the partition function by a saddle point
approximation. For chiral theories in the 4-d sense, the integrand in
(\ref{partition function:original expression}) is not invariant under
this $Z_2$ so that a saddle point approximation using this measure does not
respect the above $Z_2$ symmetry of eigenvalue distribution.
It is suggested in \cite{Amariti:2011jp} that we make the integrand to be
symmetric under the $Z_2$ symmetry by dropping the irrelevant $Z_2$ odd part,
and then carry out the saddle point approximation. Although the partition function
$Z$ after $\sigma$ integration is just the same, we call the latter expression
as $Z_{sym}$. $Z_{sym}$ is given by
\begin{eqnarray}
\hspace*{-1.4cm}Z_{sym}&=&\frac{1}{2}\int\left(\prod_\textrm{Cartan}
\frac{d\sigma}{2\pi}\right)\exp\left[\frac{ik}{4\pi}\sum_i
\left(-2{\sigma_{1,i}}^2+{\sigma_{2,i}}^2+{\sigma_{3,i}}^2\right)
+\frac{i}{4\pi}\sum_{a,b}\sum_{i,j}\Lambda_{ab}\sigma_{a,i}\sigma_{b,j}\right]\nonumber\\
\hspace*{-1.4cm}&&\left(\prod_a\prod_{i,j,i\neq j}2\sinh\frac{\sigma_{a,i}-\sigma_{a,j}}{2}\right)\left\{\exp\left[-\sum_a\sum_i\Delta^m_a\sigma_{a,i}
+\sum_I\sum_{i,j}\ell\left(1-\Delta^I+i\frac{\sigma_{1,i}-\sigma_{2,j}}{2\pi}\right)+cyclic\right]
\right.\nonumber\\
\hspace*{-1.4cm}&&~~\left.+\exp\left[\sum_a\sum_i\Delta^m_a\sigma_{a,i}+\sum_I\sum_{i,j}
\ell\left(1-\Delta^I-i\frac{\sigma_{1,i}-\sigma_{2,j}}{2\pi}\right)+cyclic\right]\right\}.
\end{eqnarray}
For $\sigma_{1,i}$, the saddle point equation is given by
\begin{eqnarray}
0&=&-\frac{\partial F_{sym}}{\partial \sigma_{1,i}}\nonumber\\
&=&-\frac{ik}{\pi}\sigma_{1,i}+\frac{i}{2\pi}\sum_b\sum_j\Lambda_{1b}\sigma_{b,j}+\sum_{j\neq i}\coth\frac{\sigma_{1,i}-\sigma_{1,j}}{2}\nonumber\\
&&+A_+\left[-\Delta^m_1-\frac{i}{2}\sum_I\sum_j\left(1-\Delta^I+i\frac{\sigma_{1,i}-\sigma_{2,j}}{2\pi}\right)\cot\pi\left(1-\Delta^I+i\frac{\sigma_{1,i}-\sigma_{2,j}}{2\pi}\right)\right.\nonumber\\
&&~~\left.+\frac{i}{2}\sum_I\sum_j\left(1-\Delta^I+i\frac{\sigma_{3,j}-\sigma_{1,i}}{2\pi}\right)\cot\pi\left(1-\Delta^I+i\frac{\sigma_{3,j}-\sigma_{1,i}}{2\pi}\right)\right]\nonumber\\
&&+A_-\left[\Delta^m_1+\frac{i}{2}\sum_I\sum_j\left(1-\Delta^I-i\frac{\sigma_{1,i}-\sigma_{2,j}}{2\pi}\right)\cot\pi\left(1-\Delta^I-i\frac{\sigma_{1,i}-\sigma_{2,j}}{2\pi}\right)\right.\nonumber\\
&&~~\left.-\frac{i}{2}\sum_I\sum_j\left(1-\Delta^I-i\frac{\sigma_{3,j}-\sigma_{1,i}}{2\pi}\right)\cot\pi\left(1-\Delta^I-i\frac{\sigma_{3,j}-\sigma_{1,i}}{2\pi}\right)\right]
\end{eqnarray}
where
\begin{eqnarray}
A_\pm&=&\exp\left[\mp\sum_a\sum_i\Delta^m_a\sigma_{a,i}+\sum_I\sum_{i,j}\ell\left(1-\Delta^I\pm i\frac{\sigma_{1,i}-\sigma_{2,j}}{2\pi}\right)+cyclic\right]\bigg/\nonumber\\
&&\left\{\exp\left[-\sum_a\sum_i\Delta^m_a\sigma_{a,i}+\sum_I\sum_{i,j}\ell\left(1-\Delta^I+i\frac{\sigma_{1,i}-\sigma_{2,j}}{2\pi}\right)+cyclic\right]\right.\nonumber\\
&&~~\left.+\exp\left[\sum_a\sum_i\Delta^m_a\sigma_{a,i}+\sum_I\sum_{i,j}\ell\left(1-\Delta^I-\frac{\sigma_{1,i}-\sigma_{2,j}}{2\pi}\right)+cyclic\right]\right\}.\nonumber\\
\end{eqnarray}
$\sigma_{2,i}$ and $\sigma_{3,i}$ have similar saddle point equations.
From the symmetries of these equations, we expect that the
solution of these equations satisfies the following properties:
\begin{itemize}
\item The eigenvalue distribution is $Z_2$ invariant.
\item The eigenvalues for $\sigma_2$ and $\sigma_3$ are the same.
\end{itemize}
In adopting the method of saddle point approximation of
\cite{Jafferis:2011zi}, it is crucial that the long range force on
each eigenvalue should vanish. Here the long range force is the
force appearing when $\sigma_{a,i}-\sigma_{b,j}$ is very large. This
can be obtained by approximating $\coth
(\sigma_{a,i}-\sigma_{b,j})\sim {\rm sgn} Re(\sigma_{a,i}-\sigma_{b,j})$.
One can easily check with $Z=Z_{sym}$ and the gauge choice (\ref{DeltaI})
that the long range force vanishes. Since
$\sum_j\sigma_j=0$ and
$\sum_{i,j}\ell\left(1-\Delta^I+i\frac{\sigma_{1,i}-\sigma_{2,j}}{2\pi}\right)
=\sum_{i,j}\ell\left(1-\Delta^I-i\frac{\sigma_{1,i}-\sigma_{2,j}}{2\pi}\right)$,
the symmetrized partition function on the saddle point is reduced to
\begin{eqnarray}
Z_{sym,saddle}&=&\exp\left[\frac{ik}{4\pi}\sum_i\left(-2\sigma_{1,i}^2+\sigma_{2,i}^2+\sigma_{3,i}^2\right)\right]\left(\prod_a\prod_{i,j,i\neq j}2\sinh\frac{\sigma_{a,i}-\sigma_{a,j}}{2}\right)\nonumber\\
&&\exp\left\{\sum_I\sum_{i,j}\left[\ell\left(1-\Delta^I+i\frac{\sigma_{1,i}-\sigma_{2,j}}{2\pi}\right)+\ell\left(1-\Delta^I-i\frac{\sigma_{1,i}-\sigma_{2,j}}{2\pi}\right)\right]+cyclic\right\}.\nonumber\\
\end{eqnarray}
This saddle point value of the partition function has exactly the
same form as that of a nonchiral theory. For the $Z_2$ invariant
eigenvalue distribution, with the absence of long-range forces on
the eigenvalues, we can adopt the ansatz
\begin{equation}
\sigma_a(x)=N^\alpha x+iy_a(x)
\end{equation}
in the large $N$ limit. The leading order contribution of each
component of the free energy is then given by
\begin{equation}
F_{ext}=-\frac{ik}{4\pi}\sum_i\left(-2\sigma_{1,i}^2+\sigma_{2,i}^2+\sigma_{3,i}^2\right)\approx\frac{k}{2\pi}N^{\alpha+1}\int_{-x_*}^{x_*}
dx\rho(x)x\left[-2y_1(x)+y_2(x)+y_3(x)\right],
\end{equation}
\begin{eqnarray}\label{F_int}
F_{int}&=&-\sum_a\sum_{i,j,i\neq j}\ln\left(2\sinh\frac{\sigma_{a,i}-\sigma_{a,j}}{2}\right)\nonumber\\
&&-\sum_I\sum_{i,j}\left[\ell\left(1-\Delta^I+i\frac{\sigma_{1,i}-\sigma_{2,j}}{2\pi}\right)+\ell\left(1-\Delta^I-i\frac{\sigma_{1,i}-\sigma_{2,j}}{2\pi}\right)\right]+cyclic\nonumber\\
%&=&-\sum_{i,j~i<j}\left\{\sum_a\left[\ln e^{\sigma_{a,j}-\sigma_{a,i}}+2\ln\left(1-e^{-(\sigma_{a,j}-\sigma_{a,i})}\right)\right]\right.\nonumber\\
%&&~~+\frac{1}{2}\sum_I\left[\ell\left(1-\Delta^I+i\frac{\sigma_{1,j}-\sigma_{2,i}}{2\pi}\right)+\ell\left(1-\Delta^I-i\frac{\sigma_{1,j}-\sigma_{2,i}}{2\pi}\right)\right.\nonumber\\
%&&~~~~\left.\left.+\ell\left(1-\Delta^I+i\frac{\sigma_{2,j}-\sigma_{1,i}}{2\pi}\right)+\ell\left(1-\Delta^I-i\frac{\sigma_{2,j}-\sigma_{1,i}}{2\pi}\right)\right]+cyclic\right\}\nonumber\\
%&&+\frac{1}{2}\sum_I\sum_i\left[\ell\left(1-\Delta^I+i\frac{\sigma_{1,i}-\sigma_{2,i}}{2\pi}\right)+\ell\left(1-\Delta^I-i\frac{\sigma_{1,i}-\sigma_{2,i}}{2\pi}\right)\right]+cyclic\nonumber\\
%&\approx&-N^2\int_{-x_*}^{x*}\rho(x)dx\int_{-x_*}^{x}\rho(x')dx'\sum_n\left\{-\frac{2}{n}\sum_ae^{-n(\sigma_a(x)-\sigma_a(x'))}\right.\nonumber\\
%&&~~+\frac{1}{2}\sum_I\left[e^{-n(\sigma_1(x)-\sigma_2(x'))}\left(\frac{\phi^I\cos(n\phi^I)}{\pi n}-\left(\frac{\sigma_1(x)-\sigma_2(x')}{\pi n}+\frac{2}{\pi n^2}\right)\sin(n\phi^I)\right)\right.\nonumber\\
%&&~~~~\left.\left.+e^{-n(\sigma_2(x)-\sigma_1(x'))}\left(\frac{\phi^I\cos(n\phi^I)}{\pi n}-\left(\frac{\sigma_2(x)-\sigma_1(x')}{\pi n}+\frac{2}{\pi n^2}\right)\sin(n\phi^I)\right)\right]+cyclic\right\}\nonumber\\
&=&N^{2-\alpha}\int_{-x_*}^{x_*}dx\rho(x)^2\sum_If^I(y_a)
\end{eqnarray}
where $x_*=\textrm{Max}(x)$ and $\rho(x)$ is the eigenvalue density
function. Detailed calculation is similar to that in
\cite{Cheon:2011vi}. The function $f^I(y_a)$ on the last line is
given by
\begin{eqnarray}
f^I(y_a)&=&2\sum_n\left[\frac{1}{n^2}-\frac{\phi^I}{2\pi n^2}\cos(n\phi^I)\cos(n(y_2-y_1))+\frac{y_2-y_1}{2\pi n^2}\sin(n\phi^I)\sin(n(y_2-y_1))\right.\nonumber\\
&&~~\left.+\frac{1}{\pi
n^3}\sin(n\phi^I)\cos(n(y_2-y_1))+cyclic\right].
\end{eqnarray}
where $\phi^I=2\pi(1-\Delta^I)$. Since the eigenvalue distributions
of $\sigma_2$ and $\sigma_3$ are the same, we can set
$y_2-y_1=y_3-y_1=y$. In this case \cite{Cheon:2011vi},
\begin{equation}
f^I(y)=\frac{\phi^I}{2\pi}\left((2\pi-\phi^I)^2-y^2\right),\qquad-2\pi<y<2\pi.
\end{equation}
In order to have nontrivial saddle point, $F_{ext}$ and $F_{int}$
should be balanced, which implies that $\alpha=\frac{1}{2}$. Now we
should determine the eigenvalue density function $\rho(x)$ and the
imaginary part of the eigenvalue difference $y(x)$. Since $\rho(x)$
is constrained by the condition $\int dx\rho(x)=1$, we write the
leading order contribution of the partition function as follows:
\begin{equation}
F=N^\frac{3}{2}\left[\frac{k}{\pi}\int_{-x_*}^{x_*}
dx\rho(x)xy(x)+\int_{-x_*}^{x_*}
dx\rho(x)^2\sum_If^I(y(x))-\frac{\mu}{2\pi}\left(\int_{-x_*}^{x_*}
dx\rho(x)-1\right)\right]
\end{equation}
with the Lagrangian multiplier term. Using the variational method,
we obtain equations
\begin{eqnarray}
\frac{k}{\pi}xy(x)+2\rho(x)\sum_If^I(y(x))-\frac{\mu}{2\pi}=0,\\
\frac{k}{\pi}\rho(x)x+\rho(x)^2\sum_I{f^I}'(y(x))=0
\end{eqnarray}
whose solution is
\begin{equation}
\rho(x)=\frac{1}{2x_*},\qquad
y(x)=\frac{kx_*}{\pi}x,\qquad\qquad-x_*\leq x\leq x_*
\end{equation}
Since we have not determined $x_*$ yet, we should find the value of
$x_*$ that extrimizes the partition function
\begin{equation}
F=N^\frac{3}{2}\left[\frac{k^2}{6\pi^2}x_*^3+\frac{\sum_I\phi^I(2\pi-\phi^I)^2}{4\pi}\frac{1}{x_*}\right].
\end{equation}
It is easy to find that
$x_*=\frac{\left[2\pi\sum_I\phi^I(2\pi-\phi^I)^2\right]^\frac{1}{4}}{\sqrt{2k}}$
extremizes $F$. Substituting this value, the partition function is
given by
\begin{equation}
F=\frac{\sqrt2}{3\pi}\left[\frac{\left(\sum_I\phi^I(2\pi-\phi^I)^2\right)^3}{2\pi}\right]^\frac{1}{4}k^\frac{1}{2}N^\frac{3}{2}
=\frac{4\sqrt2\pi}{3}\left[\sum_I\left(1-\Delta^I\right){\Delta^I}^2\right]^\frac{3}{4}k^\frac{1}{2}N^\frac{3}{2}.
\end{equation}
This partition function is a function of the trial R-charges
$\Delta^I$. Extremizing the partition function with respect to
$\Delta^I$, we obtain the expected free energy
\begin{equation}
F=\frac{16\sqrt3\pi}{27}k^\frac{1}{2}N^\frac{3}{2}.
\end{equation}
with $\Delta^I=\frac{2}{3}$. This result is exactly the same as that
of the gravity theory.

\vskip 0.5cm  \hspace*{-0.8cm}
{\bf\large Acknowledgements}
\vskip 0.2cm

\hspace*{-0.75cm} This work is supported by the BK21 program of the
Ministry of Education, Science and Technology (SK), the National
Research Foundation of Korea (NRF) Grants No. 2009-0085995 (JP),
2010-0007512 (SK, DG) and 2005-0049409 through the Center for Quantum
Spacetime (CQUeST) of Sogang University (JP, SK).
JP appreciates APCTP for its stimulating environment for research.

\end{document}